\documentstyle[mnras_cite,epsfig]{mn}

\def\gsim{~\rlap{$>$}{\lower 1.0ex\hbox{$\sim$}}}
\def\lsim{~\rlap{$<$}{\lower 1.0ex\hbox{$\sim$}}} 
\def\d{{\rm d}}

\begin{document}

\title[Self-Consistent Theory of Halo Mergers - II: CDM Power Spectra]{Self-Consistent
Theory of Halo Mergers - II: CDM Power Spectra}
\author[Andrew~J.~Benson etc.]{Andrew~J.~Benson$^1$ \\
$^1$Mail Code 130-33, California Institute of Technology,
Pasadena, CA~91125, U.S.A. (e-mail: {\tt abenson@tapir.caltech.edu})}

\maketitle

\begin{abstract}
We place additional constraints on the three parameters of the dark matter halo merger rate function recently proposed by Parkinson, Cole \& Helly by utilizing Smoluchowski's coagulation equation, which must be obeyed by any binary merging process which conserves mass. We find that the constraints from Smoluchowski's equation are degenerate, limiting to a thin plane in the three dimensional parameter space. This constraint is consistent with those obtained from fitting to N-body measures of progenitor mass functions, and provides a better match to the evolution of the overall dark matter halo mass function, particularly for the most massive halos. We demonstrate that the proposed merger rate function does not permit an exact solution of Smoluchowski's equation and, therefore, the choice of parameters must reflect a compromise between fitting various parts of the mass function. The techniques described herein are applicable to more general merger rate functions, which may permit a more accurate solution of Smoluchowski's equation. The current merger rate solutions are most probably sufficiently accurate for the vast majority of applications.
\end{abstract}


\section{Introduction}\label{sec:intro}

In current cosmological theory the mass density of the Universe is dominated by dark matter. The most successful model of structure formation is that based upon the concept of cold dark
matter (CDM). In the CDM hypothesis dark-matter particles interact only via the gravitational force. Since the initial distribution of density perturbations in these models has greatest power on small scales, the first objects to collapse and form dark-matter halos are of low mass. Larger objects form through the merging of these smaller sub-units. Consequently, the entire process of structure formation is thought to proceed in a ``bottom-up'', hierarchical manner.

Clearly then, the rate of dark-matter halo mergers is an absolutely crucial ingredient in models of galaxy and large-scale-structure formation, from sub-galactic scales to galactic and galaxy-cluster scales. The Press-Schechter (PS) formalism \cite{PS74} has long provided a simple, intuitive, and surprisingly accurate formula for the distribution of halo masses at a given redshift over a large range of mass scales and for a vast variety of initial power spectra.

An elegant paper by \scite{LacCol93}---and similar work by \scite{Bonetal91} and \scite{Bow91}---extended the work of Press and Schechter to determine the rate at which halos of a given mass merge with halos of some other mass. In addition to providing valuable physical insight, these merger rates have extraordinary practical value, having been applied to galaxy-formation models, e.g., if galaxy morphologies are determined by the merger history \cite{GotKlyKra99}; AGN activity \cite{WyiLoe03}; models for Lyman-break galaxies \cite{Koletal99}; abundances of binary supermassive black holes (SMBHs) \cite{VolHaaMad02}; rates for SMBH coalescence \cite{MilMer01}
and the resulting LISA event rate \cite{MenHaiNar01,Hae94}; the first stars \cite{SanBroKam02,ScaSchFer03}; galactic-halo substructure \cite{KamLid00,BulKraWei00,Benetal02,Som02,StiWidFri01}; halo angular momenta \cite{Vivetal02} and concentrations \cite{Wecetal02}; galaxy clustering \cite{Peretal03}; particle acceleration in clusters \cite{GabBla03}; and formation-redshift distributions for galaxies and clusters and thus their distributions in size, temperature, luminosity, mass, and velocity \cite{Veretal01,VerHaiSpe02}.

However, as first noted by Lacey \& Cole (1993; see their Section~3.1) and as we demonstrated in Benson, Kamionkowski \& Hassani (2005; hereafter Paper~I]) these merger rates are fundamentally flawed. The extended-Press-Schechter (ePS) formulae for merger rates are mathematically self-inconsistent, providing {\it two} different results for the same merger rate, as was also pointed out by \scite{SanBroKam02}, which are increasingly discrepant for larger mass ratios.  This ambiguity will be particularly important for, e.g., understanding galactic substructure and for SMBH-merger rates \cite{Erickcek06}. Even the smaller numerical inconsistency for mergers of nearly equal mass may be exponentially enhanced during repeated application of the formula while constructing merger trees to high redshift. Moreover, the ambiguity calls into question the entire formalism, even when the two possibilities seem to give similar answers quantitatively\footnote{Extended Press-Schechter theory discusses the trajectories of points in the primordial density field as the smoothing scale is reduced. It is the association of such trajectories with halo masses, which is not necessarily well-defined (see, for example, \protect\scite{porc02} who find that only 40\% of proto-halo regions contain peaks of the density field), that leads to these problems with the derived merger rates.}. Recently, \scite{neidek08b} have described an alternative derivation of merger rates from ePS theory and an implementation which appears to be self-consistent. Our goal here, however, is not to find a merger rate kernel which reproduces the results of ePS since that theory does not reproduce results found in N-body simulations. Instead, we aim to find a merger kernel which agrees well with N-body results and correctly evolves mass functions designed to fit the halo mass function measured from N-body simulations.

In Paper~I, we discussed the mathematical requirements of a self-consistent theory of halo mergers. As recognized already \cite{SilWhi78,CavMen92,She95,ShePit97}, the merger process is described by the Smoluchowski coagulation equation. Indeed, several authors have employed the Smoluchowski equation in precisely this way \cite{cava91,CavMen92,cava93,menci02}. This equation simply says that
the rate at which the abundance of halos of a given mass changes is determined by the difference between the rate for creation of such halos by mergers of lower-mass halos and the rate for destruction of such halos by mergers with other halos. The correct expression for the merger rate must be one that yields the correct rate of evolution of the halo abundance when inserted into the coagulation equation. The problem is thus to find a merger rate, or ``kernel,'' that is consistent with the evolution of the halo abundance, either the Press-Schechter abundance or one of its more recent N-body--inspired variants \cite{STMF,J2000MF}.

The apparent simplicity of the mathematical problem is in fact quite deceptive. The Smoluchowski coagulation equation is in fact an infinite set of coupled nonlinear differential equations. The equation appears in a variety of areas of science---e.g., aerosol physics, phase separation in liquid mixtures, polymerization, star-formation theory \cite{AllBas95,SilTak79}, planetesimals
\cite{Wet90,MalGoo01,Lee00}, chemical engineering, biology, and population genetics---so there is a vast but untidy literature on the subject (although see \pcite{leyvraz} for an illuminating review). It has been studied a little by pure and applied mathematicians \cite{Ald99}. Still, solutions to the coagulation equation are poorly understood. Furthermore, there is virtually no literature on the problem we face: i.e., how to find a merger kernel that, when inserted
into the coagulation equation, yields the desired halo mass distribution and its evolution as a solution.

In Paper~I we solved Smoluchowski's equation numerically subject to two additional physically-motivated constraints (or regularization conditions). First, the merger kernel should be positive for all masses\footnote{A negative merger rate could be considered to describe a spontaneous fission process, but such processes should have a different dependence on halo abundances and so require the inclusion of additional terms in Smoluchowski's equation. We ignore such processes in this work.} and, second, the merger kernel should be a smooth function of its arguments. The second regularization condition was imposed by minimizing the second derivatives of the kernel with respect to the two arguments. This allowed us to find unique solutions to Smoluchowski's equation for several power-law power spectra. These solutions were in reasonable agreement with the limited N-body data available for comparison.

We have attempted to extend the work carried out in Paper~I to the physically interesting case of a CDM power spectrum, using a much improved calculation able to span a wider dynamic range of halo masses. While we have been able to obtain solutions to Smoluchowski's equation for such power spectra we find that they \emph{do not} correspond to the merger rates that occur in N-body simulations of structure formation, which we consider to provide the ``correct'' solution for our purposes. Specifically, while our solutions are correct solutions of the Smoluchowski equation, they do not reproduce the progenitor mass functions of halos as measured from N-body simulations. (In fact, they typically perform worse than the standard extended Press-Schechter theory.) This is, perhaps, not surprising as our results are controlled by the regularization conditions applied. We required solutions to be smooth, which seems reasonable, but our minimization is forced to pick the \emph{smoothest} solution in order to find a unique answer. It seems, therefore, that the Universe uses a smooth merger rate, but not the smoothest. Trials with alternative regularization conditions have failed to find any more successful approach.

Until such time as improved physical understanding of the physics governing the merger kernel is uncovered we have adopted a different approach to finding a suitable merger kernel. Recently, several authors have used the Millennium Simulation to provide constraints on either the merger rate \cite{fma07} or the progenitor mass functions of dark matter halos \cite{neidek08}. In particular, Parkinson, Cole \& Helly (2008; hereafter PCH08) utilized progenitor mass functions to constrain the parameters of a function used to modify the standard extended Press-Schechter merger rate function and thereby constructed a merger tree binary split algorithm which accurately matched those progenitor mass functions. This approach is successful, but is limited by the accuracy and extent of the N-body data. In this paper, we demonstrate how we can use Smoluchowski's equation to provide an additional constraint on such algorithms. This approach is powerful for the following reasons:
\begin{enumerate}
\item The PCH08 merger tree algorithm is a binary split algorithm and so should obey Smoluchowski's equation;
\item Constraining the three free parameters of the PCH08 algorithm is much easier than attempting to numerically invert Smoluchowski's equation subject to complicated regularization conditions;
\item Smoluchowski's equation applies over all halo masses, while even the Millennium Simulation has a limited dynamical range of masses that it can probe.
\end{enumerate}
The disadvantage of this approach is that it assumes a particular functional form for the merger kernel (with free parameters that are to be fit). There is no guarantee that this functional form will permit a solution to Smoluchowski's equation for any values of its free parameters and, in fact, we will show that it does not. Nevertheless, we can find the best-fit to the Smoluchowski equation. Further free parameters could be introduced as constraints, of course, which should result in improved agreement with Smoluchowski's equation.

The remainder of this paper is arranged as follows. In \S\ref{sec:methods} we describe the Smoluchowski equation as used in this work and, in particular, how it is applied to the construction of merger trees. In \S\ref{sec:results} we present the results of fitting the parameters of the PCH08 model to N-body progenitor mass functions and to the Smoluchowski equation and examine how these influence the evolution of the dark-matter halo mass function. Finally, in \S\ref{sec:disc} we examine the implications of these results.

\section{Smoluchowski's Equation and Merger Trees}\label{sec:methods}

In this section we will describe the techniques used in this work and the specific implementations used to construct progenitor mass functions and evolving dark-matter halo mass functions.

Smoluchowski's equation describes the changing distribution of ``masses''\footnote{The equation applies to any additive quantity conserved through the coagulation process.} of objects growing through coagulation. Given a distribution, $n(M;t)$, of object masses $M$ at time $t$, the Smoluchowski equation gives the rate of change of this distribution:
\begin{eqnarray}
\dot{n}(M;t) & = & \int_0^{M/2} Q(M-M^\prime,M^\prime;t) n(M-M^\prime;t) n(M^\prime;t) \d M^\prime \nonumber \\
 & &  - \int_0^\infty Q(M,M^\prime;t) n(M;t) n(M^\prime;t) \d M^\prime,
\label{eq:SmolM}
\end{eqnarray}
where $Q(M_1,M_2;t)$ encodes the merger rate between objects of mass $M_1$ and $M_2$ at time $t$. It is, therefore, symmetric in its arguments, i.e. $Q(M_1,M_2;t)=Q(M_2,M_1;t)$. The first term in eqn.~(\ref{eq:SmolM}) represents creation events, while the second represents destruction events. Only three forms for $Q(M_1,M_2;t)$ are known to permit analytic solutions of Smoluchowski's equation: $Q(M_1,M_2;t)=k$ (where $k$ is a constant), $Q(M_1,M_2;t)=k(M_1+M_2)$ and $Q(M_1,M_2;t)=kM_1M_2$. In this work, the objects we consider are dark matter halos, and the mass is the total mass of those halos. The second analytic solution is of particular interest as it corresponds to the solution for dark matter halos in a Universe with a $P(k)=k^n$ power spectrum when $n=0$. The other two analytic solutions are not cosmologically interesting.

It is convenient to work with scaled variables for mass and time. As shown in Paper~I, the natural time variable is $\tau = -\ln[\delta_{\rm c})(t)]$ where $\delta_{\rm c}(t)$ is the extrapolated linear theory overdensity required for halo collapse in the spherical top-hat collapse model (as appears in the Press-Schechter expression for the halo mass
function). We also choose to express masses in units of the characterstic mass scale $M_*$ (defined such that $\sigma(M_*[t])=\delta_{\rm c}(t)\equiv \exp(-\tau)$ where $\sigma^2(M)$ is the fractional mass variance of the linear density field extrapolated to ${\mathcal Z}=0$ (we use ${\mathcal Z}$ to indicate redshift so as to distinguish from the mass variable $z$). With these choices, the Smoluchowski equation becomes:
\begin{eqnarray}
y(z;\tau) & = & \int_0^{z/2} q(z-z^\prime,z^\prime;\tau) n(z-z^\prime;\tau) n(z^\prime;\tau) \d z^\prime \nonumber \\
 & &  - \int_0^\infty q(z,z^\prime;\tau) n(z;\tau) n(z^\prime;\tau) \d z^\prime,
\label{eq:Smolz}
\end{eqnarray}
where $z=M/M_{\rm *}(t)$, $n(z)$ is the distribution of halo masses, $y(z)=\d n(z)/\d\tau$ and $q(z_1,z_2;\tau)$ is the merger rate function in these new variables. For the specific case of power-law power-spectra (which are, of course, scale-free) there is no explicit time-dependence with this choice of units and we can write
\begin{eqnarray}
y(z) & = & \int_0^{z/2} q(z-z^\prime,z^\prime) n(z-z^\prime) n(z^\prime) \d z^\prime \nonumber \\
 & &  - \int_0^\infty q(z,z^\prime) n(z) n(z^\prime) \d z^\prime.
\end{eqnarray}
Thus, a single solution, valid at all times can be found for power-law power-spectra. For CDM power-spectra, which are not scale free, the merger kernel can in principle depend explicitly on time (although the choice of scaled variables will minimize this dependence). In principle, therefore, we could use Smoluchowski's equation at each point in time ($\tau$) to provide constraints on the merger rate function. We retain the variable choices $z$ and $\tau$ as we expect the solutions $q(z_1,z_2;\tau)$ to be only slowly changing with $\tau$ in these variables, since the CDM power-spectrum is close to power-law over a wide range of masses. In practice, we will use only the $\tau=\tau_0$ (i.e. present day) Smoluchowski equation, although our methods could be easily applied to other redshifts if required.

It is well known that the Press-Schechter mass function is not a good description of the mass functions found in N-body simulations of CDM structure formation \cite{STMF,J2000MF}. Our methods are applicable to any mass function $n(z)$ and so we can replace the Press-Schechter mass function with a fitting formula such as that from \scite{STMF}:
\begin{eqnarray}
n_{\rm SMT}(z;\tau) & = & \sqrt{2\over \pi} A_{\rm SMT} \alpha(z) {x^\prime(z,\tau) \over z^2} \exp[ -x^{\prime 2}(z,\tau)/2] \nonumber \\
 & & \times  (1+1/x^{\prime 2q_{\rm SMT}}),
\label{eqn:zSMTabundance}
\end{eqnarray}
where $x^\prime(z,\tau) = \sqrt{a_{\rm SMT}} x(z,\tau)$, $x(z,\tau)=\exp(-\tau)/\sigma(z)$, $a_{\rm SMT}=0.707$, $q_{\rm SMT}=0.3$ and $A_{\rm SMT}(\approx 0.3222)$ is chosen such that the mass density in halos equals the mean density of the Universe.

Smoluchowski's equation involves the time derivative of this function, which is given by
\begin{equation}
y_{\rm SMT}(z,\tau) = n_{\rm SMT}(z,\tau) \left[ x^{\prime 2}(z,\tau) + {2q_{\rm SMT} \over 1 + x^{\prime 2q_{\rm SMT}}(z,\tau)} - 1 \right].
\label{eq:ySMT}
\end{equation}

As shown in Paper~I, the standard extended Press-Schechter theory predicts a merger kernel
\begin{eqnarray}
     Q_{\rm ePS}(M_1,M_2;t) &=& {M_2^2 \over \rho_0 \sigma_{\rm f}} {
     \sigma_2 \over M_{\rm f}} \left|{\dot \delta_c \over \delta_c}
     \right| \left| {\d \ln \sigma_f \over \d \ln M_{\rm f}} \right|
     \left| {\d \ln \sigma_2 \over \d \ln M_2} \right|^{-1} \nonumber \\
     & \times & {1
     \over (1-\sigma_{\rm f}^2/\sigma_1^2)^{3/2}} \nonumber \\
     &\times& \exp \left[ - {\delta_c^2 \over 2} \left( {1\over
     \sigma_{\rm f}^2} - {1\over \sigma_2^2} - {1\over \sigma_1^2} \right)
     \right],
\label{eq:qeps}
\end{eqnarray}
where we have adopted the notation $\sigma_2 = \sigma(M_2)$, etc. This is, of course, asymmetric in the two mass arguments. In scaled variables, this becomes
\begin{eqnarray}
     q_{\rm ePS}(z_1,z_2;\tau) &=& {z_2^2 \over \sigma_{\rm f}} {
     \sigma_2 \over z_{\rm f}}
     \left| {\d \ln \sigma_f \over \d \ln z_{\rm f}} \right|
     \left| {\d \ln \sigma_2 \over \d \ln z_2} \right|^{-1} \nonumber \\
     & \times & {1
     \over (1-\sigma_{\rm f}^2/\sigma_1^2)^{3/2}} \nonumber \\
     &\times& \exp \left[ - {{\rm e}^{-2\tau} \over 2} \left( {1\over
     \sigma_{\rm f}^2} - {1\over \sigma_2^2} - {1\over \sigma_1^2} \right)
     \right],
\label{eq:qepssc}
\end{eqnarray}
where, as shown in Paper~I, the mean cosmic density, $\rho_0$, is removed through an appropriate choice of units.

When computing the mass variance, $\sigma(M)$, we use the same power spectrum as used to generate the initial conditions for the Millennium Simulation since we will fit to conditional mass functions measured from that simulation. This power spectrum is integrated under a spherical top-hat real-space window function to obtain $\sigma(M)$.

\subsection{Application to Merger Trees}\label{sec:mergertrees}

A primary goal of this study is to provide accurate merger rates for use in the construction of dark matter halo merger trees. ``Accurate'' here is defined to mean that the merger trees so constructed should produce the correct distribution of progenitor halo masses at earlier times and, consequently, produce the correct evolution of the halo mass function. Ideally, this evolution should remain precise even when trees are constructed over large fractions of cosmic history. The need for such accurate trees is highlighted by the work of \scite{BensonReion06} (see their Fig.~9), who were forced to restart their calculations of reionization at periodic intervals in redshift due to accumulating inaccuracies in their merger trees. Constructing merger trees also provides a means of comparing our results for the merger rate with measurements from N-body simulations. In this Section, therefore, we will describe how we construct merger trees.

\subsubsection{Merger Tree Construction Algorithm}

We adopt the algorithm described by \scite{cole00} and PCH08 to construct merger trees. This algorithm considers binary splits of the merger tree with a correction for accretion of mass in halos below the mass resolution of the tree and uses adaptive timesteps to ensure that the probability of non-binary mergers remains small. As \scite{cole00} note, their algorithm performs well due to the subtle choice of drawing a progenitor mass from the lower half of the ePS progenitor mass distribution function (i.e. for a parent halo of mass $M_0$ they use only the $<M_0/2$ region of the progenitor mass function when computing the probability of a binary split). It is interesting to realize that this choice effectively symmetrizes the Smoluchowski merger kernel as
\begin{equation}
 q_{\rm ePS,sym}(z_1,z_2) = \left\{ \begin{array}{ll}
                                     q_{\rm ePS}(z_1,z_2) & \hbox{if } z_1 \le (z_1+z_2)/2 \\
                                     q_{\rm ePS}(z_2,z_1) & \hbox{if } z_1 > (z_1+z_2)/2. \\
                                    \end{array}
 \right.
\end{equation}
Other algorithms have been proposed for constructing merger trees \cite{KW93,cole94,SK99}--we select the \scite{cole00} because it allows only binary mergers and is therefore consistent with a treatment via the Smoluchowski equation. We refer the reader to \scite{cole00} and PCH08 (their Appendix~A) for a full description of the merger tree construction algorithm, but address the aspects of the implementation unique to the present work below.

\scite{cole00} compute two quantities, $F(z,\tau)$ and $P(z,\tau)$, which give the fraction of mass gained through accretion and the probability of a binary split respectively for a halo of mass $z$ at time $\tau$ and during some time period $\delta \tau$. In terms of quantities used in this work, the parameters $F(z,\tau)$ and $P(z,\tau)$ are given by:
\begin{equation}
F(z,\tau) = \delta \tau \int_0^{z_{\rm min}} z^\prime R(z^\prime;\tau) \d z^\prime,
\label{eq:F}
\end{equation}
and
\begin{equation}
P(z,\tau) = \delta \tau \int_{z_{\rm min}}^{z/2} R(z^\prime;\tau) \d z^\prime,
\label{eq:P}
\end{equation}
where
\begin{equation}
R(z^\prime) = {n(z^\prime;\tau) n(z-z^\prime;\tau) q(z^\prime,z-z^\prime;\tau) \over n(z;\tau)},
\end{equation}
$z_{\rm min}=M_{\rm min}/M_*(\tau)$ and $M_{\rm min}$ is the mass resolution of the merger tree. We have expressed this rate in terms of the merger kernel in Smoluchowski's equation. PCH08 give expressions for $F$ and $P$ in terms of the ePS expression for the mean number of progenitors of mass $M_1$ expected in a small timestep. The two ways of expressing these functions are equivalent---we use the version depending on the merger kernel to make clear the connection to Smoluchowski's equation as used in this work. We adaptively choose timesteps $\delta \tau$ to ensure that $F\ll 1$ and $P\ll 1$, to keep the possibility of multiple fragmentation small and to ensure that the halo mass can change by only a small amount due to accretion in any given timestep.

Using the Press-Schechter mass function for $n(z;\tau)$ and the ePS merger rate $q_{\rm ePS}(z_1,z_2;\tau)$ (eqn.~\ref{eq:qepssc}) recovers the expressions given by \scite{cole00} for $F$ and $P$. Equations~(\ref{eq:F}) and  (\ref{eq:P}) are more general however, allowing any mass function $n(z;\tau)$ and merger rate function $q(z_1,z_2;\tau)$ to be used. In particular, PCH08, propose that $R(z^\prime;\tau)$ be modified by multiplying by a function
\begin{equation}
G(\sigma_1/\sigma_{\rm f},\delta_{\rm c}(\tau)/\sigma_{\rm f}) = G_0 [\sigma_1/\sigma_{\rm f}]^{\gamma_1} [\delta_{\rm c}(\tau)/\sigma_{\rm f}]^{\gamma_2}
\end{equation}
where $G_0$, $\gamma_1$ and $\gamma_2$ are free parameters. In terms of the Smoluchowski equation we can similarly multiply the ePS merger kernel by the same function $G(\sigma_1/\sigma_{\rm f},\delta_{\rm c}(\tau)/\sigma_{\rm f})$. We will call this modified merger kernel $q^\prime_{\rm ePS}$ and label such kernels as $(G_0,\gamma_1,\gamma_2)$. However, we must also account for the fact that the PCH08 algorithm implicitly utilize the Press-Schechter mass function, while we actually want to reproduce the evolution of the \scite{STMF} mass function and must therefore use that when solving Smoluchowski's equation. Appendix~\ref{app:PS2ST} describes how to correctly implement the Smoluchowski kernel in this case.

\subsubsection{Building Progenitor Mass Functions}

Utilizing the above algorithm we can, given a halo of mass $z_{\rm f}$ at the present day, construct a merger tree back to some earlier time. By constructing a large number of such trees and averaging over their progenitors we can construct an estimate of the progenitor mass function at that earlier time. Specifically, we wish to construct progenitor mass function to compare to those determined by \scite{CHFP08} from the Millennium Simulation. \scite{CHFP08} estimated progenitor mass functions of ${\mathcal Z}=0$ halos by locating all ${\mathcal Z}=0$ halos with a mass within a factor $\sqrt{2}$ of some mass $M_{\rm f}$ and then finding all progenitors of those halos at redshifts $0.5$, $1$, $2$ and $4$. To compare to these N-body data\footnote{Available from {\tt http://star-www.dur.ac.uk/$\sim$cole/merger\_trees/MS\_data/}.} we first produce a sample of halo masses $M$ between $M_{\rm f}/\sqrt{2}$ and $\sqrt{2}M_{\rm f}$ drawn from the \scite{STMF} mass function. We then construct a merger tree for each such halo back to $z=4$ and cumulate the progenitor halo masses to estimate a mean conditional mass function.

We construct enough trees that our comparison is limited by noise in the N-body results and not by the limited number of trees constructed. We then determine a $\chi^2$ statistic using
\begin{equation}
\chi^2 = \sum \left[ {f_{\rm cmf}^{\rm MS}(M_1|M_{\rm f}) - f_{\rm cmf}^{\rm MC}(M_1,M_{\rm f}) \over \sigma_{\rm cmf}^{\rm MS}(M_1|M_{\rm f})} \right]^2,
\end{equation}
where $f_{\rm cmf}(M_1|M_{\rm f})$ is the conditional mass function of halos of mass $M_1$ which are progenitors of a halo of mass $M_{\rm f}$ at ${\mathcal Z}=0$, and superscripts refer to the Millennium Simulation (MS) and Monte-Carlo (MC) merger trees constructed using our algorithm. The values of $\sigma_{\rm cmf}^{\rm MS}(M_1|M_{\rm f})$ used are determined from the number of progenitor halos in the Millennium Simulation assuming Poisson statistics. This process is repeated for different values of the parameters $G_0$, $\gamma_1$ and $\gamma_2$.

\subsubsection{Evolving the Halo Mass Function}

In order to study the evolution of the halo mass function, $n_{\rm SMT}(z;\tau)$, we begin by drawing a sample of dark matter halo masses at random from that mass function at some initial time $\tau_1$ corresponding to ${\mathcal Z}=0$. We use a quasi-random method (specifically a Sobol sequence; \pcite{numrec} section~7.7) to produce a non-uniform sample of masses while minimizing fluctuations in the mass function due to random sampling. We impose a minimum and maximum number of halos to simulate from each decade of halo mass to ensure good sampling of the entire mass range of interest, and span a large enough range of initial masses to ensure that our results are unaffected by the necessarily limited range of masses probed.

We then apply the \scite{cole00} method to evolve these halos backwards in time until a time $\tau_2$ is reached. Summing the progenitor halo masses of all of the $\tau_1$ halos allows us to construct the halo mass function at $\tau_2$. This is then compared to the expected mass function $n_{\rm SMT}(z;\tau_2)$.

\subsubsection{Constraints from Smoluchowski's Equation}

Given a merger kernel, $q^\prime_{\rm ePS}$, we can, for any combination of parameters $(G_0,\gamma_1,\gamma_2)$, evaluate the creation and destruction integrals in Smoluchowski's equation and thereby determine $y(z)$ for that merger kernel. In general, a merger kernel will not provide a precise solution to Smoluchowski's equation. We can measure the ability of any given merger kernel to solve Smoluchowski's equation by evaluating the mean absolute difference between the resulting $y(z)$ and that found by direct differentiation of the mass function with respect to time (which is, of course, the required solution to give the correct evolution of the mass function). Therefore, we compute
\begin{equation}
\langle d \rangle = {1\over N}\sum_{i=1}^N \left| {y(z_i)-y_{\rm SMT}(z_i) \over n_{\rm SMT}(z_i)} \right|,
\label{eq:smolfit}
\end{equation}
where $y_{\rm SMT}(z)$ is the rate of change of the \scite{STMF} mass function, and we average the absolute error over $N=16$ different masses, $z_i$, equally spaced in $\log z$ between $z=10^{-5}$ and $z=10^3$. We have chosen to weight by $1/n_{\rm SMT}(z)$ so that we are comparing the rate of change of the mass function per halo. The choice of how to judge any merger kernel's degree of success in solving Smoluchowski's equation is somewhat subjective---the above choice ensures that we require the solution to be most accurate where the timescale for change in the mass function is most rapid (i.e. for the most massive halos).

\section{Results}\label{sec:results}

\begin{figure*}
\begin{tabular}{cc}
  \psfig{file=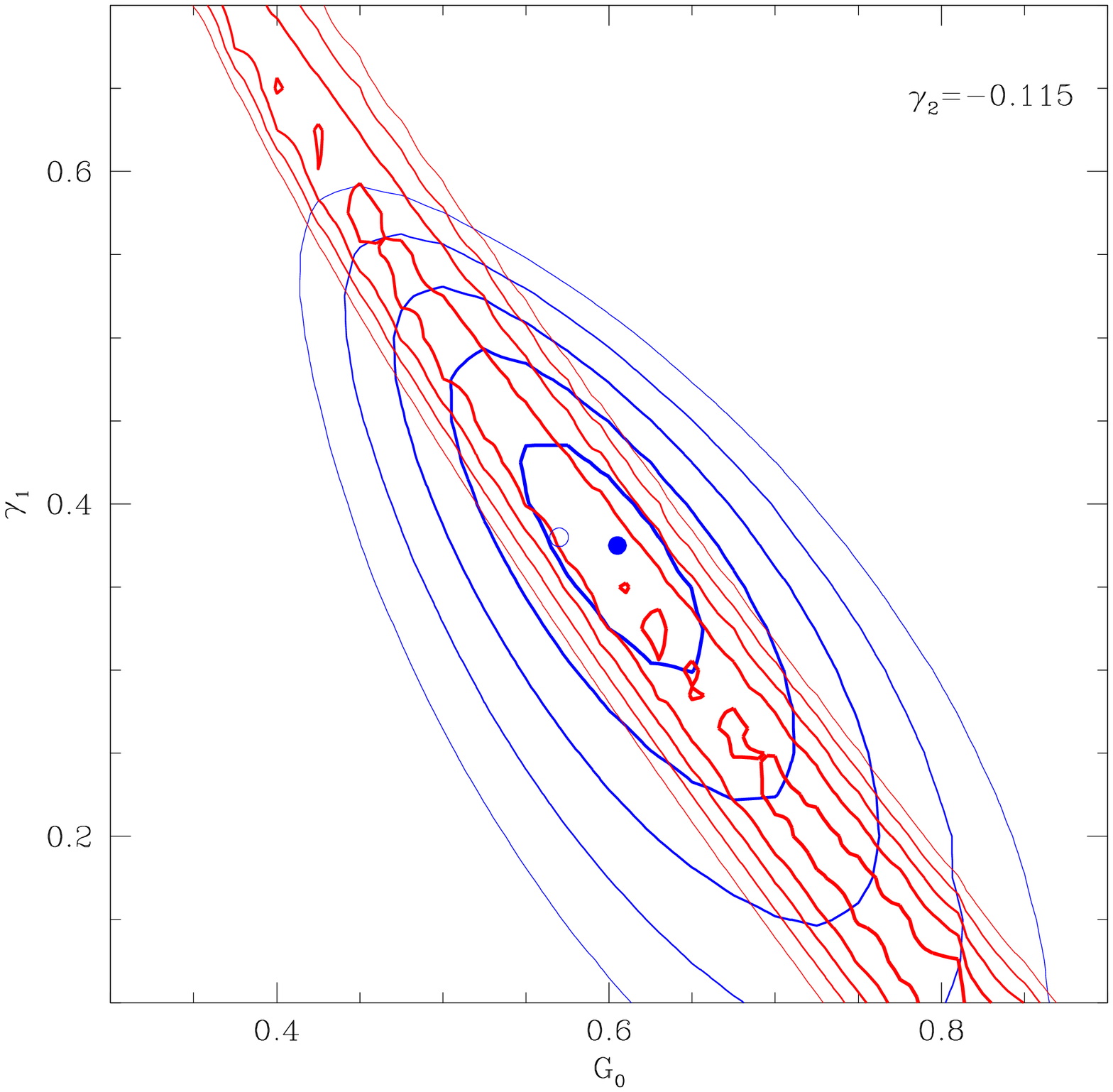,width=80mm}
& \psfig{file=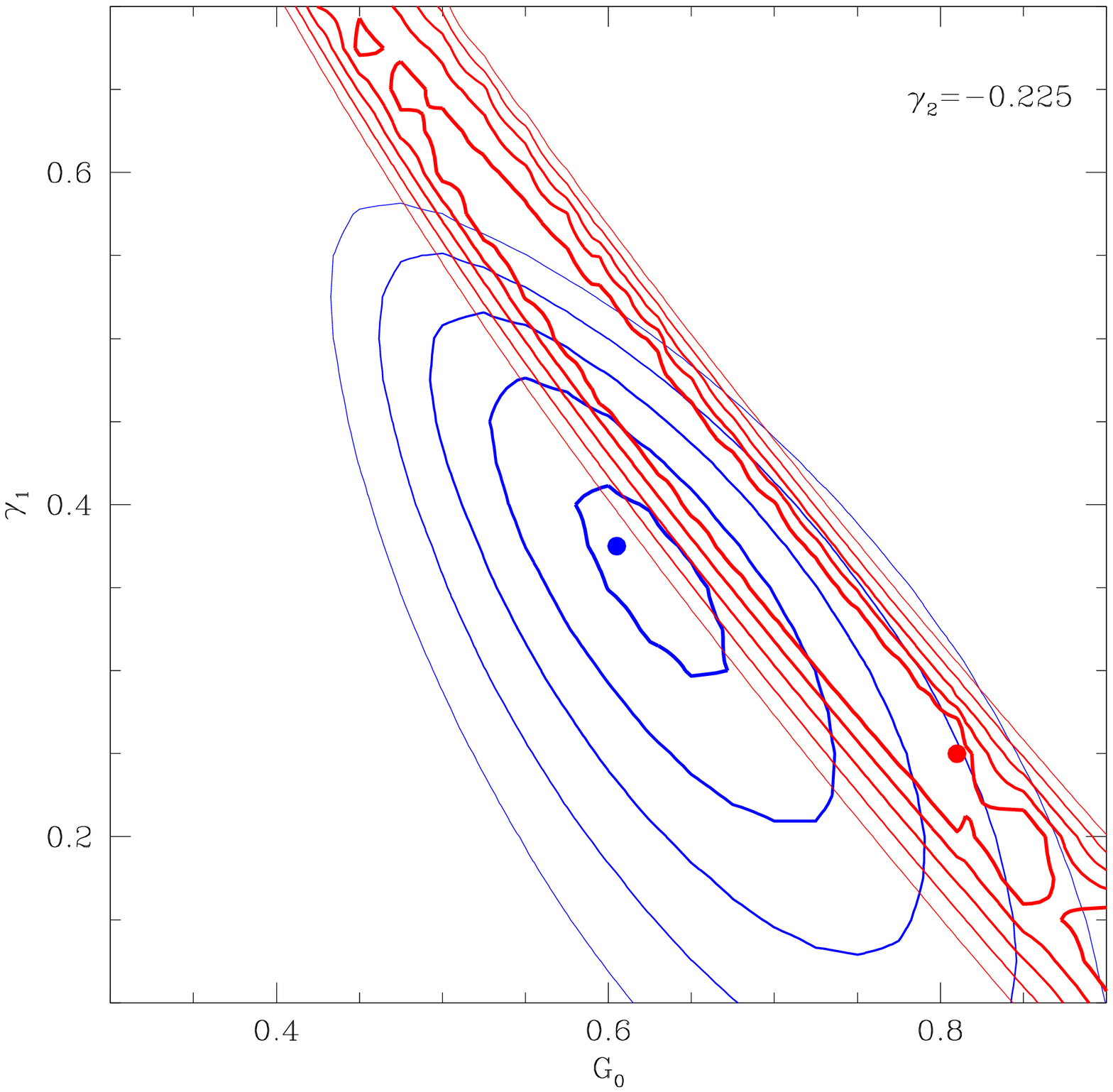,width=80mm}
\end{tabular}
\caption{Constraints in the $G_0$-$\gamma_1$ parameter plane for two values of $\gamma_2$ ($\gamma_2=-0.125$ in the left-hand panel, corresponding to the best fit to Millennium Simulation conditional mass functions; $\gamma_2=-0.225$ in the right-hand panel corresponding to the best solution of Smoluchowski's equation). The solid blue dot shows the best-fit values obtained by fitting to progenitor mass functions measured from the Millennium Simulation, while blue contours show the constraints obtained from this dataset. The red point shows the best-fit values obtained by fitting to the Smoluchowski equation, while the red contours map locii of constant mean error between $y(z)$ and the fitted value as defined in eqn.~(\ref{eq:smolfit}). The constraints shown correspond to a slice through the 3-D parameter space at constant $\gamma_2$. The open blue circle shows the constraint obtained by PCH08 using the same Millennium Simulation data set. This differs slightly from our equivalent constraint due to our choice to utilize the measurement errors in the progenitor mass functions in our fit. Nevertheless, the two are quite similar as expected. (Note that PCH08 found a best fitting value of $\gamma_2=-0.01$.)}
\label{fig:constraints}
\end{figure*}

We find best-fit parameters of $(G_0,\gamma_1,\gamma_2)=(0.605,0.375,-0.115)$ and $(0.775,0.275,-0.225)$ for fits to the Millennium Simulation conditional mass functions and Smoluchowski's equation respectively. Our best-fit parameters for matching conditional mass functions differ slightly from those obtained by PCH08 due to our choice to weight our goodness-of-fit measure by the inverse Poisson errors on the N-body conditional mass functions. As expected, both constraints favour $G_0<1$ (which reduces the overall merger rate) and $\gamma_1>0$ (which boosts the ratio of low mass to high mass progenitors)---PCH08 discuss the reasons for these expectations. Figure~\ref{fig:constraints} shows constraints in the $(G_0,\gamma_1)$ plane (i.e. on slices of fixed $\gamma_2$ in the 3-D parameter space), for both fitting to conditional mass functions (blue contours) and Smoluchowski's equation (red contours). In the left and right-hand panels we show results corresponding to the best-fitting values of $\gamma_2$ for conditional mass functions and Smoluchowski's equation respectively.

The conditional mass functions provide a good constraint on $(G_0,\gamma_1)$ with a degree of degeneracy between the parameters. Smoluchowski's equation shows a much larger degeneracy between these two parameters and, in fact, favours regions which lie in a plane in the full 3-D parameter space. This plane is, however, thin, thereby strongly ruling out large regions of the parameter space. Remarkably, the degeneracy seen in fitting Smoluchowski's equation is similar to that from fitting conditional mass functions. More importantly, the constraints from both methods intersect, permitting solutions which both match the conditional mass functions and are reasonable solutions to Smoluchowski's equation. This is an important point, as it implies that N-body merger trees are consistent with a binary merger hypothesis.

We could, in principle, combine these two constraints to obtain an overall best-fitting model. Such a procedure is, however, somewhat arbitrary. Firstly, while we have a statistically meaningful (in the sense that we can use it to assign relative probabilities to models) measure of goodness-of-fit for the fits to conditional mass functions, our goodness-of-fit to Smoluchowski's equation is not statistically meaningful. Any combination of the two constraints therefore requires some amount of judgment as to their relative importance. Furthermore, the $(G_0,\gamma_1,\gamma_2)$ modification to extended Press-Schechter merger rates does not give a ``good-fit'' to conditional mass functions (it is certainly a much better fit than unmodified extended Press-Schechter, but clear systematic differences from the N-body data remain) and does not precisely solve Smoluchowski's equation.

It is, however, obvious that any choice of a ``best-fit'' model to both constraints must lie somewhere in the plane defined by Smoluchowski's equation between the red and blue dots in Fig.~\ref{fig:constraints}. We will illustrate below the differences in results when these two best-fit values are obtained---any combined best-fit should show intermediate behaviour.

\begin{figure}
\psfig{file=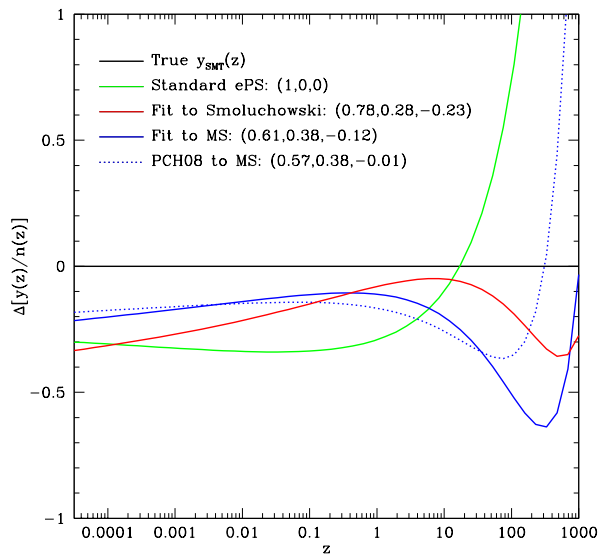,width=80mm}
\caption{The rate of change of the halo mass function per halo, $y(z)/n(z)$, relative to that obtained by direct differentiation of the \protect\scite{STMF} mass function (the horizontal black line indicates the true rate of change expressed in this way). Coloured lines show this quantity computed using the Smoluchowski equation using various merger kernels, i.e.  using $q^\prime_{\rm ePS}$ with different parameters $(G_0,\gamma_1,\gamma_2)$. The green line shows the result of using the standard extended Press-Schechter kernel, $(1,0,0)$, while the red line shows the results for the merger kernel which best matches the true rate of change per halo as judged by eqn.~(\protect\ref{eq:smolfit}). The solid blue line shows results for the merger kernel that best fits progenitor mass functions in the Millennium Simulation. The dotted blue line shows the same using the fit parameters of PCH08. For all calculations, we use the \protect\scite{STMF} mass function in Smoluchowski's equation.}
\label{fig:y_fits}
\end{figure}

It is instructive to examine how well our various models solve Smoluchowski's equation. Figure~\ref{fig:y_fits} shows the difference between the rate of change of the mass function per halo as a function of halo mass determined from our merger kernels and the value required for a correct solution of Smoluchowski's equation. The standard extended Press-Schechter merger kernel (green line) fails significantly over the entire range of masses plotted (particularly so at the massive end). The blue lines indicate results using merger kernels constrained to match the conditional mass functions from the Millennium Simulation. Over the range where these conditional mass functions provide a good measure of the merger kernel (approximately $0.001<z<10$) the fitted merger kernels provide a significantly better solution to Smoluchowski's equation than does the unmodified extended Press-Schechter kernel. Note also that our fit (solid blue line) and that of PCH08 (dotted blue line) are very similar over this range. At $z\gsim 10$ the Millennium Simulation conditional mass functions provide only weak constraints on the merger kernel and, consequently, the kernels fitted to these functions provide worse solutions to Smoluchowski's equation in this regime (although still better than the original extended Press-Schechter kernel). Finally, the red line shows results from the kernel which provides the best solution to Smoluchowski's equation. ``Best'' here is in the sense defined by eqn.~(\ref{eq:smolfit}), which tries to solve Smoluchowski's equation most accurately where the rate of change of the mass function per halo is largest. Not surprisingly, therefore, this gives the most accurate solution for massive halos ($z\gsim 1$), but actually gives a worse solution to Smoluchowski's equation at low masses than do kernels fit to conditional mass functions. This simply reflects the relative importance given to different mass ranges by each constraint and the fact that the particular functional form of the merger kernel chosen does not permit a precise solution to Smoluchowski's equation, thereby forcing a compromise solution to be adopted.

\begin{figure*}
\psfig{file=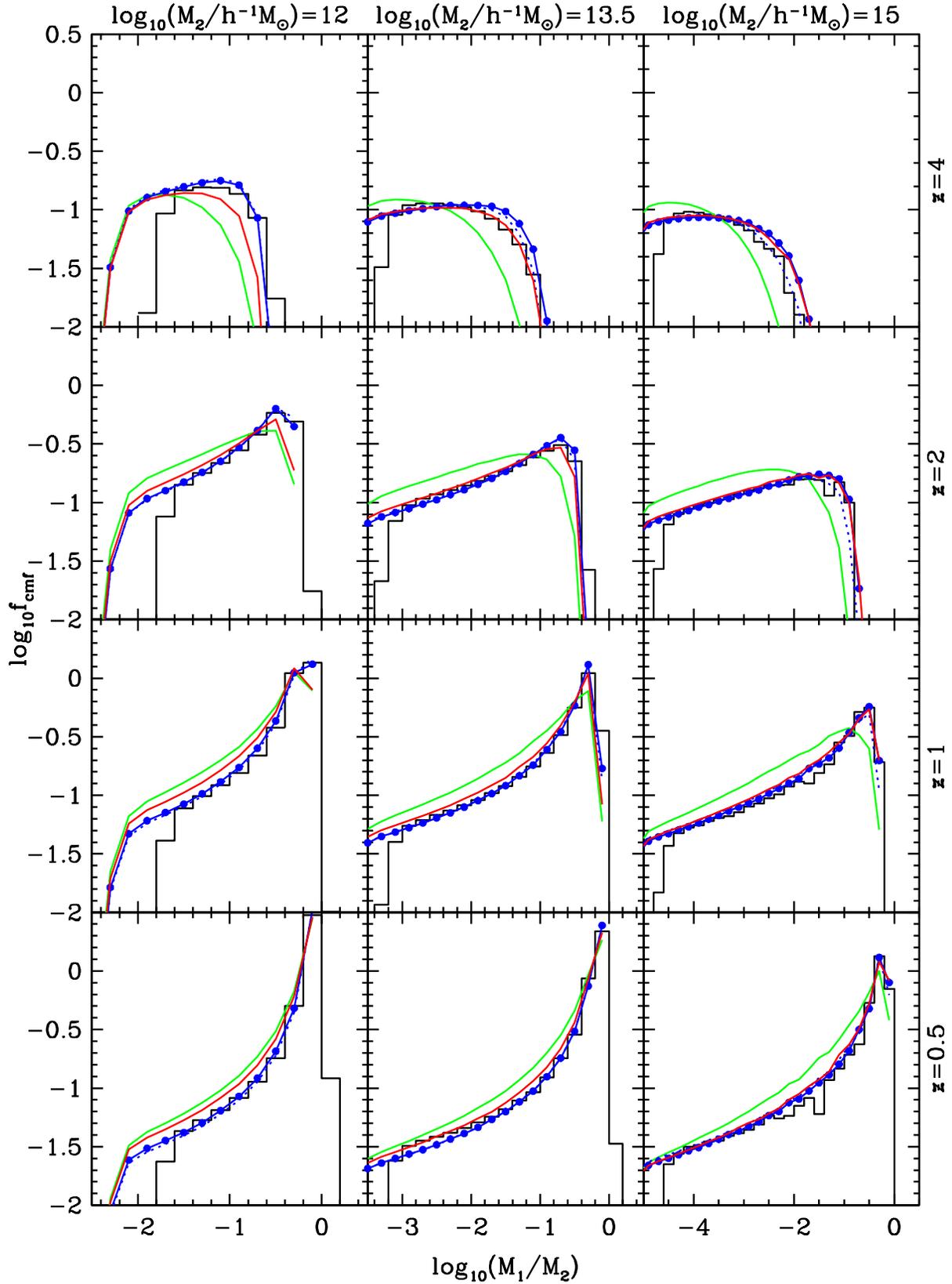,bbllx=7mm,bblly=15mm,bburx=190mm,bbury=270mm,clip=,width=160mm}
\caption{Comparison of progenitor mass functions measured from the Millennium Simulation (black histograms; taken from \protect\scite{CHFP08}) and those constructed using merger trees for a variety of merger kernels, $(G_0,\gamma_1,\gamma_2)$. The green lines shows results for the standard extended Press-Schechter merger kernel, $(1,0,0)$, while the red line shows results using the merger kernel which best solves the Smoluchowski equation. The solid blue curve is the best fit to the Millennium Simulation found in this work, while the dotted blue line uses the best fit from PCH08. For all calculations, we use the \protect\scite{STMF} mass function in Smoluchowski's equation.}
\label{fig:progens}
\end{figure*}

Figure~\ref{fig:progens} (which has the same format as Figure~1 of PCH08) shows conditional mass functions from the Millennium Simulation (black histograms) for three different final halo mass ranges and for four different redshifts. Overplotted are conditional mass functions estimated from merger trees built using different merger kernels. The failure of the unmodified extended Press-Schechter kernel (green lines) is readily apparent, while blue lines---which use kernels constrained by fitting to these same conditional mass functions in this work (solid lines) and PCH08 (dotted lines)---are vastly improved matches as expected. The red line indicates progenitor mass functions obtained using the merger kernel which best solves Smoluchowski's equation. This is clearly intermediate in success between an unmodified extended Press-Schechter kernel and kernels constrained to match the conditional mass functions, as may be expected. (There are cases where it performs significantly better however, for example, for the ${\mathcal Z}=4$ conditional mass function of $10^{13.5}h^{-1}M_\odot$ halos.)

\begin{figure*}
\begin{tabular}{cc}
\psfig{file=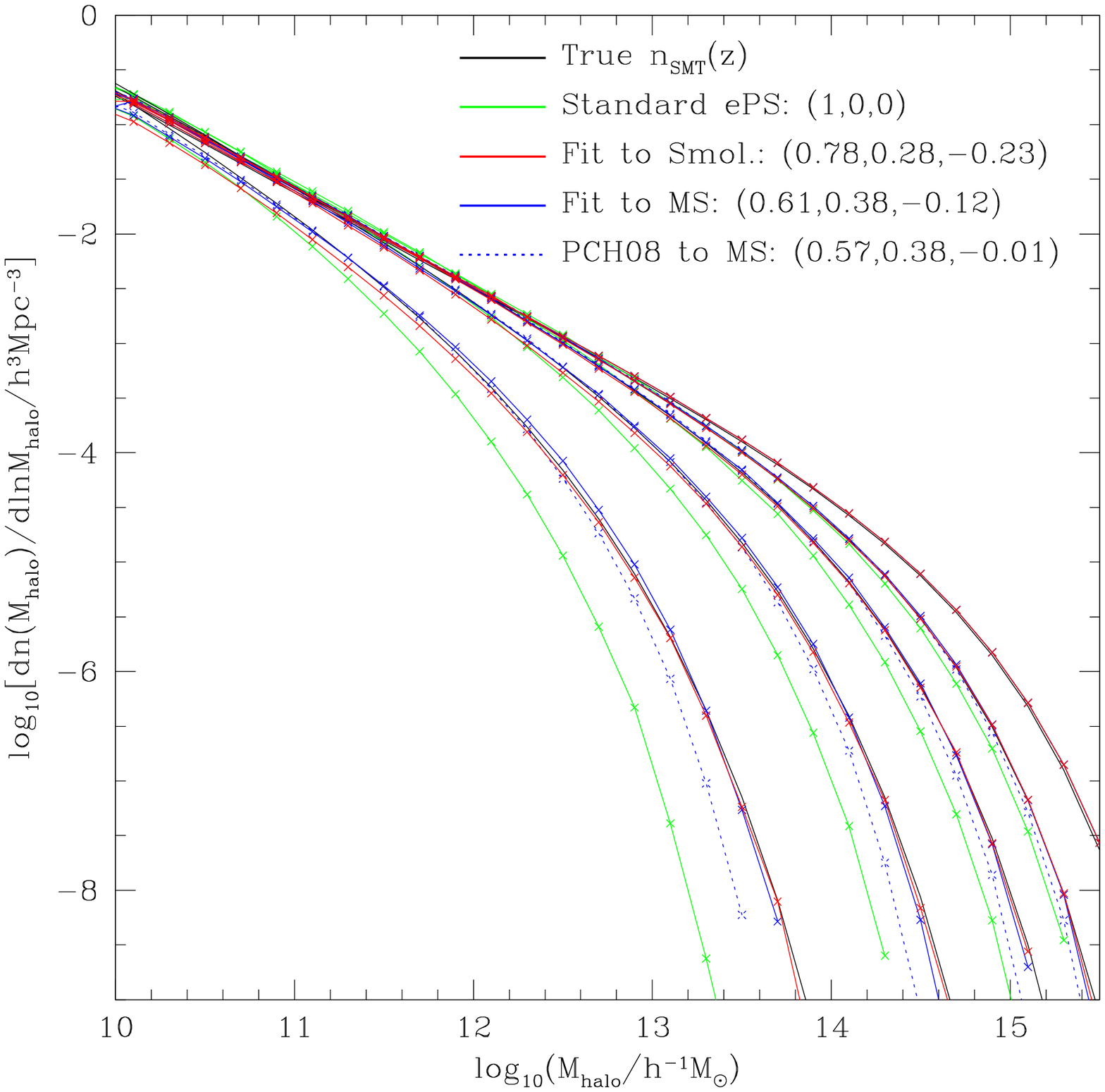,width=80mm} &
\psfig{file=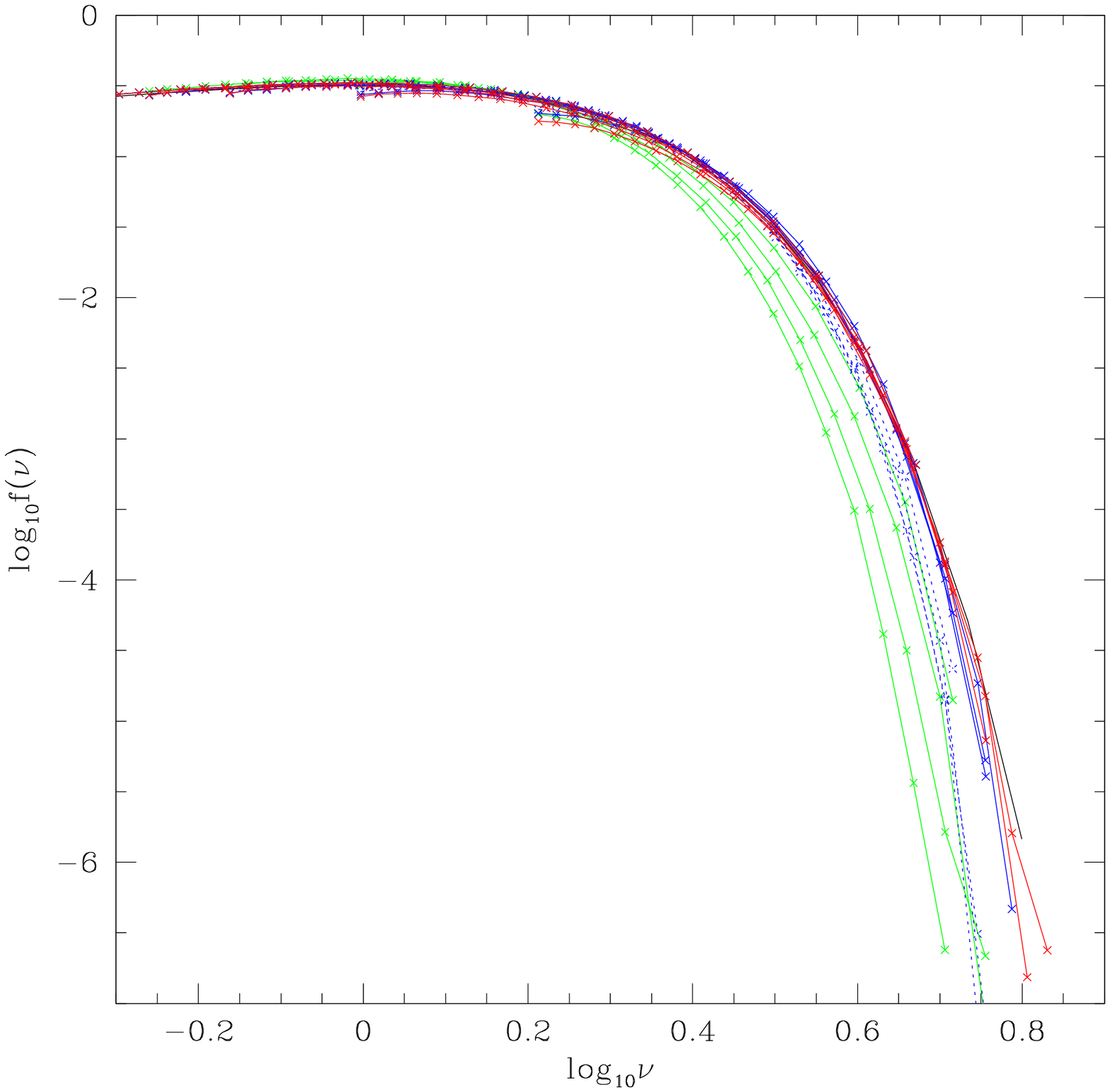,width=80mm}
\end{tabular}
\caption{\emph{Left-hand panel:} The dark matter halo mass function shown at ${\mathcal Z}=0$, $0.5$, 1, 2 and 4 (from right to left). The analytic \protect\scite{STMF} mass function is shown by black lines. Coloured lines show the result of evolving the mass function from ${\mathcal Z}=0$ to higher redshifts using merger trees with different merger kernels, $(G_0,\gamma_1,\gamma_2)$. The green lines show results for the standard extended Press-Schechter merger kernel, $(1,0,0)$, while the red lines show results using the merger kernel which best solves the Smoluchowski equation. The solid blue lines are for the the best fit to the Millennium Simulation conditional mass functions found in this work, while the dotted blue lines use the best fit to conditional mass functions from PCH08.}
\label{fig:MF}
\end{figure*}

Finally, in Fig.~\ref{fig:MF} we show total (i.e. not conditional) mass functions of halos from $z=0$ to $z=4$. The left hand panel shows this function in physical units, while the right hand panel shows the fraction of mass in halos with $\nu=\delta_{\rm c}(z)/\sigma(M)$ per logarithmic interval of $\nu$ (c.f. Figure~4 of PCH08. In this latter form the \scite{STMF} mass function is redshift independent. We probe to very high masses and very low abundances to illustrate the importance of high-accuracy merger kernels when considering rare objects. The unmodified extended Press-Schechter merger kernel performs poorly, quickly resulting in a mass function shifted to low masses. The kernel with parameters identified by PCH08 performs much better, but also begins to underpredict the abundance of the most mass halos at high-redshift. The kernel using parameters obtained in this work by fitting conditional mass functions performs extremely well, remaining remarkably close to the expected \scite{STMF} mass function out to ${\mathcal Z}=4$, although statistically significant differences can be detected. The kernel with parameters selected to best solve Smoluchowski's equation (red line) also performs remarkably well. In particular, it is the most successful at match the evolution of the most massive halos in the mass function, as would be expected from Fig.~\ref{fig:y_fits}. It performs somewhat worse than the kernel fit to conditional mass functions at low masses as also expected.

\section{Discussion}\label{sec:disc}

We have described how Smoluchowski's equation, which governs any mass-conserving binary coagulation process, may be used to provide constraints on halo merger rates. These constraints are complementary to those obtained by fitting to conditional mass functions from N-body simulations since they span a wide dynamical range of halo masses. Using these methods, and the modified extended Press-Schechter algorithm described by PCH08, we have identified merger kernels which best-fit conditional mass functions from the Millennium Simulation and which best solve Smoluchowski's equation. While the functional form of the PCH08 merger rate does not permit an exact solution of Smoluchowski's equation, we are able to find solutions which greatly improve the accuracy of merger trees. Using these best-fit solutions we are able to evolve the dark matter halo mass function from ${\mathcal Z}=0$ to ${\mathcal Z}=4$ with remarkably high accuracy, particularly for the most massive objects.

Ideally, we would identify a functional form for the merger kernel which permits a precise solution to Smoluchowski's equation, while simultaneously producing progenitor masses functions consistent with the available N-body data. In fact, a perfect solution should agree with N-body data when compared using any statistic (e.g. distributions of most massive or second most massive progenitors). In reality, we do not know of such a function and, as mentioned in \S\ref{sec:intro}, only a handful of analytic solutions to Smoluchowski's equation are known. In that case, the search for the ``best'' kernel requires making some decision about what are the most important statistics to fit and accepting that some degree of compromise in matching them is unavoidable. We believe that the most important statistics to match are the evolution of the overall mass function (which is extremely well constrained from N-body simulations) and progenitor mass functions. When considering the process of structure and galaxy formation it is these statistics which control, to first order, the number of galaxies able to form at a given point in cosmic history and how those galaxies were formed.

Practically, these modified merger rates should prove extremely useful in constructing high-accuracy merger trees for use in studies of structure formation, reionization and galaxy formation. If even greater accuracy is required, a different functional form for the merger kernel must be adopted\footnote{The functional form of PCH08's modification of the merger kernel can be considered to be the first-order terms in a Taylor expansion of $\ln G$---adding higher-order terms would be straightforward.} and the techniques described in this work applied to constrain its parameters. As noted in \S\ref{sec:intro} only a handful of analytic solutions to Smoluchowski's equation are known, none of which provide useful solutions for dark matter halo merger rates in cold dark matter Universes. Nevertheless, a sufficiently general parametrization of the merger kernel should permit arbitrarily accurate solutions to Smoluchowski's equation. If used in a binary split merger tree algorithm such a kernel should allow for an arbitrarily accurate evolution of the halo mass function (limited only by the accuracy of the merger tree construction algorithm).

This work does not provide any further physical insight into dark matter halo merger rates---at least not directly. One might hope that physical insight into the ``micro-physics'' (to make an analogy with other areas in which Smoluchowski's equation is applied, e.g. polymer growth) of dark matter halo merging might point towards a functional form for the kernel. Until such insight is gained, the methods described herein provide a practical method for rapid construction of high-accuracy merger trees.

\section*{Acknowledgments}

AJB acknowledges the support of the Gordon and Betty Moore Foundation. This work has benefitted from conversations with numerous people. In particular, we would like to thank Shaun Cole, Steve Furlanetto, Dan Grin, Marc Kamionkowski, Hannah Parkinson, Cristiano Porciani and Tristan Smith for valuable discussions.

\appendix
\onecolumn

\section{Relation Between Merger Tree Split Probabilities and Smoluchowski Equation Merger Kernel}\label{app:PS2ST}

Smoluchowski's equation is normally employed to take an existing distribution of masses and evolve it forwards in time subject to a specified merger kernel. Both creation and destruction of halos must be considered in this case. However, we can also apply Smoluchowski's equation to the reverse process, taking a distribution of halos of specified mass and evolving them backwards in time. In this case, we have a series of mass-conserving binary splits reminiscent of merger trees. The split probability is related to the creation term in Smoluchowski's equation.

PCH08 describe a merger tree binary split algorithm which utilizes a modification of the extended Press-Schechter split probability distribution. Here, we wish to cast that same algorithm in the terms used in Smoluchowski's equation.

Assuming a \scite{STMF} mass function, Smoluchowski's equation is
\begin{equation}
 y_{\rm SMT}(z) = {1\over 2} \int_0^z n_{\rm SMT}(z^\prime) n_{\rm SMT}(z-z^\prime) q(z^\prime,z-z^\prime) {\rm d}z^\prime - \int_0^\infty n_{\rm SMT}(z) n_{\rm SMT}(z^\prime) q(z,z^\prime) {\rm d}z^\prime,
\end{equation}
where $n_{\rm SMT}(z)$ is the \scite{STMF} mass function and $y_{\rm SMT}(z)$ is the rate of change of that mass function.

Applying Smoluchowski's equation in reverse, we find that in a binary split algorithm, the split probability distribution for a single halo of mass $z$ to have a progenitor of mass $z^\prime$ (and, therefore, a second progenitor of mass $z-z^\prime$) is given by
\begin{equation}
P_{\rm split}(z^\prime;z) = {n_{\rm SMT}(z^\prime) n_{\rm SMT}(z-z^\prime) \over n_{\rm SMT}(z)} q(z^\prime,z-z^\prime).
\end{equation}

The PCH08 split probability function is that derived from the Press-Schechter mass function and extended Press-Schechter techniques:
\begin{equation}
P_{\rm split}(z^\prime;z) = {n_{\rm PS}(z^\prime) n_{\rm PS}(z-z^\prime) \over n_{\rm PS}(z)} q^\prime_{\rm ePS}(z^\prime,z-z^\prime),
\end{equation}
where $q^\prime_{\rm ePS}(z_1,z_{\rm f}-z_1)=q_{\rm ePS}(z_1,z_{\rm f}-z_1)G(\sigma_1/\sigma_{\rm f},\delta_{\rm f}/\sigma_{\rm f})$ and $G(\sigma_1/\sigma_{\rm f},\delta_{\rm f}/\sigma_{\rm f})$ is PCH08's multiplicative modifier of merging rates.

From this, we can deduce that the rate at which systems of mass $z_1$ and $z_2$ merge together is just
\begin{equation}
 R(z_1,z_2) = n_{\rm SMT}(z_1+z_2) P_{\rm split}(z_1;z_1+z_2) = {n_{\rm SMT}(z_1+z_2) \over n_{\rm PS}(z_1+z_2)}n_{\rm PS}(z_1) n_{\rm PS}(z_2) q^\prime_{\rm ePS}(z_1,z_2)
\end{equation}

Therefore,
\begin{equation}
 y_{\rm SMT}(z) = {n_{\rm SMT}(z) \over n_{\rm PS}(z)} {1\over 2} \int_0^z n_{\rm PS}(z^\prime) n_{\rm PS}(z-z^\prime) q^\prime_{\rm ePS}(z^\prime,z-z^\prime) {\rm d}z^\prime - \int_0^\infty
{n_{\rm SMT}(z+z^\prime) \over n_{\rm PS}(z+z^\prime)} n_{\rm PS}(z) n_{\rm PS}(z^\prime) q^\prime_{\rm ePS}(z,z^\prime) {\rm d}z^\prime,
\end{equation}
or, in terms of the rate of change per halo,
\begin{equation}
 {y_{\rm SMT}(z) \over n_{\rm SMT}(z)} = {1 \over n_{\rm PS}(z)} \left[ {1\over 2} \int_0^z n_{\rm PS}(z^\prime) n_{\rm PS}(z-z^\prime) q^\prime_{\rm ePS}(z^\prime,z-z^\prime) {\rm d}z^\prime - \int_0^\infty
{n_{\rm PS}(z)n_{\rm SMT}(z+z^\prime) \over n_{\rm SMT}(z) n_{\rm PS}(z+z^\prime)} n_{\rm PS}(z) n_{\rm PS}(z^\prime) q^\prime_{\rm ePS}(z,z^\prime) {\rm d}z^\prime \right].
\end{equation}
This is the form of Smoluchowski's equation that we utilize throughout this work.

\end{document}